# Helium Detection in Technical Materials

Andrew K. Gillespie [a]*, Cuikun Lin [a], Django Jones [a], and R. V. Duncan [a]

[a] *Department of Physics and Astronomy, Texas Tech University, Lubbock Texas 79409, USA*
*Author to whom correspondence should be addressed: a.gillespie@ttu.edu

**Abstract**

Materials used to study nuclear fusion can retain atmospheric helium unless pretreated before an experiment. Understanding helium outgassing is important for accurate diagnostics in experiments surrounding nuclear fusion. The presence of helium is often cited as the primary evidence that a nuclear reaction has occurred, so it is imperative that known sources of helium are mitigated prior to proceeding with novel nuclear experiments. It is also necessary to ensure hermiticity when transferring gas aliquots from an experiment to a mass spectrometer. In this article, we present studies of detecting helium leak rates in systems used in novel nuclear experiments. We also present studies of helium retention in materials subjected to various heating profiles and atmospheric concentrations. Without pretreatment, stainless-steel 316 retains between 15 – 240 pmol of $^4$He or an areal outgassing amount of 0.07 – 1.20 pmol/cm$^2$. It also may reabsorb $^4$He from the atmosphere in time. These studies also demonstrate that it is necessary to pretreat most materials prior to performing experiments where the presence of $^4$He is being used as an indicator for novel nuclear reactions.

*Keywords:* Helium; Outgassing; Nuclear Reactions; LENR; Retention; Detector

## 1. Introduction

Fusion has become a subject of great interest in recent decades. This process involves the collision and merging of atoms, resulting in the release of an enormous amount of energy. Even small quantities of fuel possess significant inherent energy that is liberated during fusion. By bringing atomic nuclei together and overcoming their repulsive electrostatic forces, the strong nuclear force is exploited, leading to the generation of an immense amount of power. In contrast to existing nuclear power plants that rely on fission, where energy is produced through the breaking apart of heavy atoms like uranium into lighter nuclei, fusion derives energy from the merging of lightweight nuclei, primarily hydrogen. Many fusion reactor projects focus on heating hydrogen isotopes, specifically deuterium (D) and tritium (T), to create a plasma—a state of matter composed of ionized atoms and charged particles—and then initiating fusion. Unlike fission, D-T fusion generates some radiation in the form of short-lived neutrons but does not result in long-lasting radioactive waste. Moreover, fusion is considered safer than fission because it can be more readily controlled and turned off as needed. Looking towards the year 2050, it is expected that the global population will grow by 33%, accompanied by a significant increase in economic growth. Consequently, the demand for energy could increase to over five times its current levels. However, the existing energy system faces challenges in terms of environmental sustainability, economic stability, and global security. It is imperative to address these issues by meeting the escalating energy demand while transitioning to clean, affordable, and abundant energy sources. The global fusion energy market is predicted to play a substantial role in this transition. According to estimates, the market size is projected to reach $429.6 billion in 2030, with a further increase to $840.3 billion by 2040, growing at a compound annual growth rate (CAGR) of 6.9% from 2031 to 2040. Several key market players are actively involved in the development and advancement of fusion energy technologies. Some of these prominent players include: Zap Energy; First Light Fusion; General Fusion; TAE Technologies; Commonwealth Fusion; Tokamak Energy; Lockheed Martin; HyperJet Fusion; Marvel Fusion; Helion; HB11; Agni Fusion Energy, to list just a few [1]. While most fusion approaches today involve deuterium cycles, there are notable exceptions, such as TAE Technologies that involves neutron capture by boron to produce three alpha ($^4$He) nuclei. In a significant development, the Department of Energy announced in December 2022 that its scientists achieved the first-ever fusion reaction that produced more energy than it consumed—a crucial milestone on the path toward commercial fusion power. The experiment, conducted on

December 5th at the Lawrence Livermore National Laboratory in California, lasted only a few billionths of a second. However, laboratory leaders confirmed that it demonstrated the feasibility of sustained fusion power for the first time [2-6].

In most fusion reactors, the deuterium cycle comprises of four fusion reactions that can occur involving deuterium. These reactions are as follows:

$$D+D \rightarrow {}^3He+n +3.27 \text{MeV} \quad (1)$$

$$D+D \rightarrow T+p +4.03 \text{MeV} \quad (2)$$

$$D+T \rightarrow {}^4He+n +17.59 \text{MeV} \quad (3)$$

$$D+{}^3He \rightarrow {}^4He+p +18.3 \text{MeV} \quad (4)$$

One of the primary challenges in successfully commercializing fusion power lies in developing materials capable of withstanding the extreme conditions found in fusion reactors, including elevated temperatures and high fluxes of hydrogen isotopes and neutrons and alpha particles. For instance, when these helium particles bombard the tungsten wall, they aggregate to form clusters within the material. Upon reaching a critical mass, these helium clusters can displace a tungsten atom from its usual position, creating nanoscale voids within the tungsten lattice. These voids act as nuclei for helium bubbles, which can grow to significant sizes, compromising the material's durability. Additionally, these bubbles serve as traps for tritium, diminishing its availability for the fusion reaction and posing radiological risks. Furthermore, the formation of helium bubbles induces the development of a fuzzy nanostructure on the tungsten surface, which may erode into the plasma, degrade its quality, and potentially cool the fusion reaction, complicating maintenance efforts. Consequently, there is considerable interest in devising experimental and simulation methods to assess helium retention before and after prolonged exposure to radiation from neutrons, alpha particles, and other sources of varying energies [7-15]. Such endeavors are crucial for enhancing our understanding of material behavior under fusion reactor conditions and for informing the design of more resilient materials for fusion energy applications.

On the other hand, Low-energy nuclear reactions (LENRs) encompass a range of phenomena that occur under specific conditions on certain metals such as Pd, Ni, Ti, among others, in the presence of hydrogen or its other stable isotope, deuterium. Experimental research in this field typically involves mild temperature and pressure conditions, resulting in a variety of nuclear products, effects, and nuclear-scale heat. Over the past few decades, the Low Energy Nuclear Reactions (LENR) community has actively pursued investigations under different conditions to identify signatures of nuclear reactions. In 2022, the US Department of Energy (DOE) announced a funding opportunity (DE-FOA-0002784) of up to $10 million [16]. This funding aims to establish clear protocols and practices for determining whether low-energy nuclear reactions (LENR) could serve as a potentially transformative, carbon-free energy source. The initiative is part of the Advanced Research Projects Agency-Energy (ARPA-E) LENR Exploratory Topic, which seeks to overcome the research impasse in this field and drive advancements in LENR technology. Within the LENR community, researchers have been actively searching for the presence of helium-4 as a crucial signature and evidence of nuclear reactions. This is particularly important for correlating helium measurements with excess heat observations. However, several challenges arise in this pursuit. It's known that ambient air naturally contains helium at a concentration of approximately 5.2 parts per million (ppm). Consequently, helium can easily diffuse or be absorbed into experimental apparatus, leading to contamination that may significantly impact the obtained results.

In this study, various metals including palladium, stainless steel, and platinum, as well as polymers and ceramics such as nylon and PTFE underwent testing to assess helium retention characteristics before heating, after heating, or during complete evaporation of the technical materials. Of particular focus were metals like stainless steel (SS) and palladium (Pd), aiming to precisely quantify trapped helium within wire lengths through heating beyond their respective evaporation thresholds.

## 2. Experimental Design

In order to study the helium being trapped inside of the metals, an initial approach involved attempting to evaporate metal wires such as Pd, Pt, and stainless steel using an e-beam evaporator. Helium detection was carried out by evacuating the chamber with a turbo-molecular pump (TMP), with a 4He leak detector [17] serving as the fore-line pump to the TMP. However, the results revealed that the ambient helium level and pressure in this large evaporator were too unstable to facilitate the intended measurements. Although the electron beam evaporator operated correctly, it was apparent that achieving sub-picomole 4He noise levels within any standard evaporator was unlikely due to the introduction of contaminants over years of service measuring a wide range of materials.

The development of a specialized and dedicated e-beam evaporator was deemed cost-prohibitive. Thus, a more effective approach, thermal evaporator, termed the "Light Bulb Experiment," was devised. This experiment utilizes (1) a 0.009"-diameter tungsten wire with a high melting temperature of 3,422 °C as the base heater, around which the 0.009" diameter Pd wire under test is wrapped, and (2) a ConFlat (CF) 2.75" power feedthrough with 0.020'' W wire wrapped around a pre-baked ¼'' alumina ceramic tube. These components are illustrated in **Figure 1** below. The CF chamber underwent pre-baking with an input power of 600 W, employing a tungsten wire until the leak rate stabilized near 3.7E-12 torr-L/s. The exterior temperature of the chamber ranged from approximately 280 to 300°C. To set up, as illustrated in **Figure 2** Bottom Left, a copper wire and a tungsten wire were prepared. A CF 2.75" power feedthrough featuring 0.020" tungsten wire wrapping encircled a pre-baked ½" alumina ceramic tube. To facilitate mass-4 measurements through stainless steel evaporation, water cooling via an ice bath was employed. The input power was adjusted to 508 W for a duration of 3.3 hours until the baseline leak rate was achieved. In the case of SS304-Sample #1, a wire mass of 0.2157 g was utilized. The resultant integrated helium rate was approximately 2.52E-7 torr-L, detecting a helium quantity of 14.8 pmol. The leak rate and integrated leak rate data are depicted in **Figure 3** below. This particular sample underwent complete evaporation, with some deposition occurring on the chamber wall. At power levels ranging from 550 to 600W, the $Al_2O_3$ crucible reached its melting point. Notably, the melting temperature of $Al_2O_3$ is 2,072°C. In the current configuration, materials were evaporated using power levels between 500 and 550W, slightly below the melting point of $Al_2O_3$. In future iterations, $ZrO_2$ tubes will be utilized, which boast a higher melting temperature of 2715°C. This procedure was replicated with a larger mass of stainless steel, measuring 0.3228 g. The resultant integrated helium flow was approximately 3.33E-7 torr-L, with a detected helium amount of approximately 19.4 pmol. This technique has the capability to completely evaporate metals and also analyze the retention of helium within the metal.

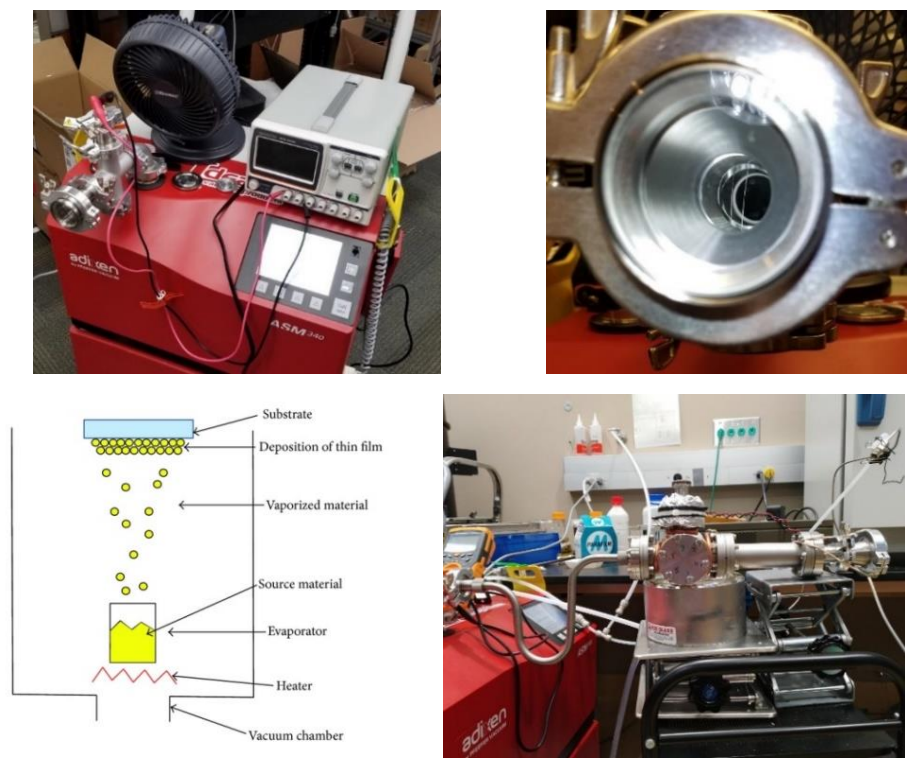

**Figure 1:** *Top Left*: The original light bulb experiment involving flanges and the ASM-340 leak detector. *Top Right*: A close-up on the end see-through flange with wires connected to input current. *Bottom Left*: The thermal evaporator with a heater, source material, and deposition substrate. *Bottom Right*: The modified light bulb experiment involving higher voltage components for evaporation of technical materials.

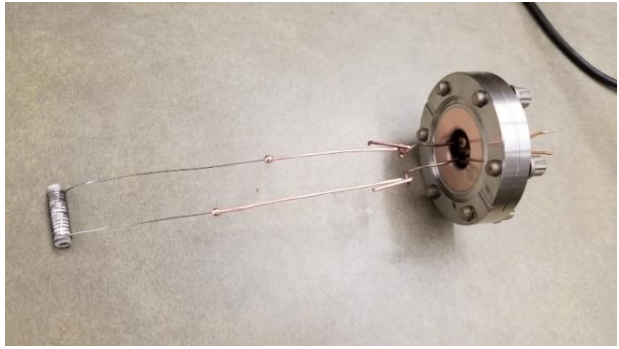 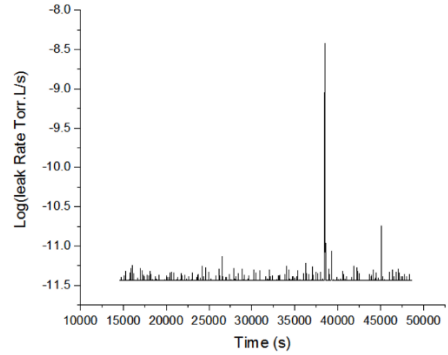

**Figure 2:** *Left*: An improved evaporator design with an alumina tube. *Right*: An example of a steady baseline leak rate near 3.7E-12 torr-L/s.

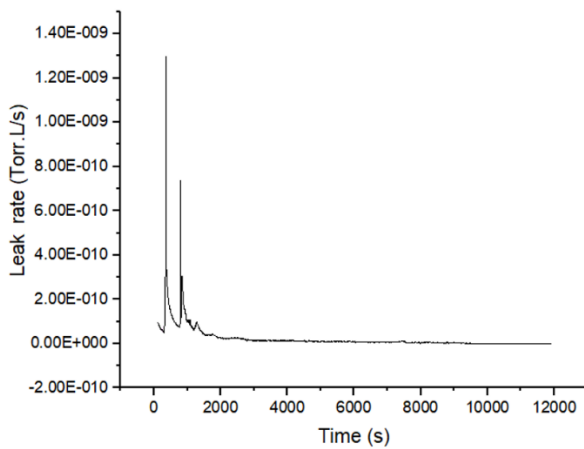 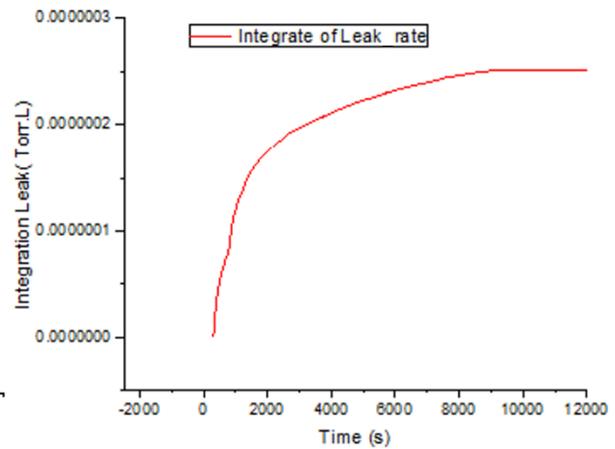

**Figure 3:** *Left*: The leak rate versus time for SS304 Sample #1. *Right*: The integrated leak rate as a function of time for sample SS304 Sample #1.

To assess the absorption amount of helium outgassed from stainless steel tube, polymer and ceramic materials, experiments were conducted employing a tube furnace (LINDENBERG/BLUE) in conjunction with a helium Leak Detector (Pfeiffer Vacuum ASM 340). The ASM 340 leak detector underwent calibration specifically for measuring $^4$He leak rates. Within the ASM 340 system, two potential test methods exist: one entails creating a hard vacuum within the system, while the other involves a sniffing test. For this series of experiments, the hard-vacuum option was selected. In the experiment aimed at detecting helium outgassing from stainless steel tubing, various lengths of stainless-steel tubing with different diameters (1/4", 3/8", 1/2") were utilized. These tubes were welded shut at one end, while the other end was connected to an ASM 340 leak detector via a valve. A J-type thermocouple was employed to measure the surface temperature of the tubing under scrutiny. Real-time temperature profiles were recorded and plotted using a LABVIEW program [18]. The ASM 340 detector was configured to monitor and display the helium outgassing rate in torr liters per second, with readings taken at one-second intervals.

To initiate the experiment, the valve between the ASM 340 leak detector and the sample was sealed, and the vacuum pump was activated to eliminate moisture, potential contaminants, and unwanted gases from the vacuum hose. Once the pump reached its baseline pressure, the valve leading to the tube sample was gradually opened to evacuate the sample. Any helium outgassing was initially observed at room temperature until the system returned to its baseline state. Subsequently, the temperature controller was engaged to heat the tube sample to 600°C within a span of 20 minutes, while simultaneously monitoring the helium outgassing rate. The temperature at which notable outgassing occurred was determined by correlating the temperature data with the leak detector data, utilizing timestamps within LABVIEW.

Upon completion of the heating process, the temperature controller was deactivated to allow the sample to cool back to room temperature. The quantity of helium outgassed in picomoles was calculated by integrating the helium leak rate over time within the peak, employing the ideal gas law.

In a similar experiment aimed at obtaining preliminary estimates of the temperature at which helium outgassing initiates and the quantity of helium released under varying temperatures, materials such as Teflon, Nylon, and MACOR were utilized. A stainless-steel tube with one open end served to contain the sample within the furnace. Initially, the stainless-steel tube was inserted into the furnace and connected to the ASM 340 leak detector. Subsequently, the tube underwent baking at 600°C for one hour while continuously connected to the helium leak detector, to eliminate any helium previously absorbed by the tubing.

After the initial baking process, the furnace was allowed to cool back to room temperature before introducing a sample for analysis. The sample was positioned inside the stainless-steel tubing, and the tubing was once again connected to the helium leak detector. The temperature controller within the furnace was programmed to achieve 600°C within a span of 10 minutes. Both the leak detector and the temperature controller were activated almost simultaneously. Throughout this preliminary experiment, the temperature increased at a rate of approximately 1°C per second.

## 3. Results and Discussion

The $^4$He leak rate was measured with respect to the time to find the temperature of helium outgassing and the amount of $^4$He picomoles outgassed. The outgassing temperature was obtained by finding the exact time stamp, the "increase in $^4$He outgassing" happened and correlating it with the temperature measurements with respect to time obtained from a LABVIEW program. The amount of picomoles outgassed are calculated using the ideal gas law.

$$PV = nRT, \quad n = \frac{PV}{RT} \tag{1}$$

Integrating $^4$He leak rate over time gives the PV term in the ideal gas law. The temperature in this calculation is used as 298 K because the helium leak rate obtained from the leak detector corresponds to this temperature. **Table 1** summarizes the $^4$He outgassing in picomoles, the outgassing temperature, and calculated values of outgassing per unit area and unit volume for tubing of different sizes. The stainless-steel tubes were pre-baked and kept exposed to atmosphere for 24 hours, 15 days and 27 days and used in the experiment to find the $^4$He outgassing rates. **Table 2** contains the resulted $^4$He picomoles outgassed for the tubes under different conditions. When performing $^4$He outgassing experiment for various materials such as Teflon, MACOR and cathodes from CF cells, 3/8" or ½" stainless steel tube with a fused end, pre-baked at 600 °C was used as the sample holder. **Table 3** summarizes the total number of picomoles of $^4$He outgassed for different samples under different conditions.

**Table 1:** Total helium amount outgassed, release temperature of out-gassing and outgassing per unit area and unit volume for different size tubing, Swagelok, Wall thickness 0.035".

| Material | Vendor | Sample | $^4$He Outgassing [pmol] | Areal $^4$He Outgassing [pmol/cm$^2$] | $^4$He Outgassing [pmol/cm$^3$] | Outgassing Temperature [°C] |
|---|---|---|---|---|---|---|
| SS 316 | Swagelok | 12-11-17 3/8"-SS | 43 | 0.29 | 0.76 | 400 |
| SS 316 | Swagelok | 02-12-18 3/8"-SS | 30 | 0.20 | 0.52 | 400 |
| SS 316 | Swagelok | 01-26-18 1/2"-SS | 240 | 1.20 | 2.20 | 400 |
| SS 316 | Swagelok | 12-11-17 3/8"-SS | 39 | 0.27 | 0.69 | 400 |
| SS 316 | Swagelok | 12-13-17 3/8"-SS | 64 | 0.44 | 1.10 | 400 |
| SS 316 | Swagelok | 03-23-18 1/4"-SS | 7.8 | 0.09 | 0.40 | 400 |
| SS 316 | Swagelok | 8-001 1/2"-SS | 64 | 0.31 | 0.57 | 400 |
| SS 316 | Swagelok | 8-002 1/2"-SS | 92 | 0.45 | 0.82 | --- |
| SS 316 | Swagelok | 8-004 1/2"-SS | 58 | 0.28 | 0.51 | 386 |
| SS 316 | Swagelok | 8-005 1/2"-SS | 36 | 0.18 | 0.32 | 386 |
| SS 316 | Swagelok | 8-006 1/2"-SS | 15 | 0.07 | 0.13 | 427 |
| SS 316 | Swagelok | 8-007 1/2"-SS | 92 | 0.44 | 0.81 | 427 |

**Table 2:** Total helium amount outgassed and release temperature of out-gassing after being exposed to atmosphere for a certain time.

| Material | Vendor | Sample | $^4$He Outgassing [pmol] | Areal $^4$He Outgassing [pmol/cm$^2$] | $^4$He Outgassing [pmol/cm$^3$] | Outgassing Temperature [°C] |
|---|---|---|---|---|---|---|
| SS 316 | Swagelok | Pre-baked, soaked in $^4$He for 24 hours | 16.21 | 0.1100 | 0.2850 | 400 |
| SS 316 | Swagelok | Pre-baked, soaked in $^4$He for 24 hours | 0.52 | 0.0036 | 0.0092 | 400 |
| SS 316 | Swagelok | Pre-baked, soaked in $^4$He for 15 days | 18.20 | 0.1250 | 0.3220 | 400 |
| SS 316 | Swagelok | Pre-baked, soaked in $^4$He for 27 days | 0.88 | 0.0060 | 0.0160 | 400 |

**Table 3:** Total helium amount outgassed and release temperature of out-gassing from different materials.

| Sample | Mass [g] | Baking Temperature [°C] | $^4$He Outgassing [pmol] | $^4$He Outgassing [pmol/g] | Outgassing Temperature [°C] |
|---|---|---|---|---|---|
| Blank Palladium Plate | --- | 600 | 2.7 | --- | --- |
| Seashore PTFE | 1.25 | 450 | 281 | 224.8 | --- |
| Large Seashore PTFE tube | 1.009 | 600 | 50 | 49.55 | Above melting temperature |
| Small Seashore PTFE tube | 1.006 | 500 | 87 | 86 | Above melting temperature |
| AWG PTFE tube | 1.006 | 500 | 48 | 47.71 | Above melting temperature |
| MACOR as delivered | 13 | 600 | 280 | 21.54 | 600 |
| MACOR Prebaked, 2 days in atmosphere | 13 | 600 | 110 | 8.46 | --- |
| MACOR Soaked in $^4$He 8 hours (25 °C) | 13 | RT | 5400 | 415.38 | RT |
| MACOR Soaked in $^4$He 8 hours (600 °C) | 13 | 600 | 3800 | 292.31 | --- |
| CF Cell Diffusion Test | --- | RT | 500 | --- | RT |
| $^4$He-soaked CF Cell 2 | --- | RT | 290000 | --- | RT |
| (4)3" Teflon Stripes, McMaster from our lab, 23.5 cm$^2$ | 7.061 | 600 | 7100 | 1005.5 | 380 |
| (4) Nylon Nuts | 0.191 | 400 | 1.6 | 8.38 | No sign of outgassing |
| Nylon Screws | 0.184 | 400 | 3.6 | 19.57 | No sign of outgassing |
| PTFE TWTT-22C | 0.66 | 550 | 9.9 | 15 | 360 |
| PTFE Plate washer | 3.857 | 550 | 19 | 4.93 | 360 |
| Nylon Nuts | 0.807 | 400 | 0.8 | 0.99 | No sign of sudden outgassing |
| Nylon Screws | 0.921 | 400 | 0.44 | 0.48 | No sign of sudden outgassing |
| Pd #2 wire D = 0.009", L = 12" | 0.15 | | 21.3 | 142 | Evaporation |
| Pd #1 wire, D = 0.009", L = 12" | 0.15 | | 3.83 | 25.5 | Evaporation |
| SS304 Capillary | 0.32 | | 19.4 | 60.6 | Evaporation |

*PTFE melting temperature: 327 °C
*Nylon melting temperature: 220 °C; decomposition temperature: 307 °C
*RT is room temperature around 23 °C

As indicated in the tables, stainless steel, nylon, Teflon, MACOR, and similar technical materials exhibit a notable capacity to retain significant amounts of $^4$He from the ambient atmosphere. Consequently, it is imperative to subject these materials to pretreatment before their utilization in experiments where $^4$He serves as an indicator for novel nuclear reactions. Further experiments were conducted to discern whether the detected mass-4 arises from $D_2$ gas or $^4$He. During the initial light bulb experiments, substantial quantities of Mass-2 were detected using the ASM-340 leak detector during the bake-out of a 0.009" diameter Pd wire. The average Mass-4 to Mass-2 ratio was calculated to be 182 ppm, closely resembling the natural abundance ratio of deuterium to protium. In a separate experiment, a 6"-long, 0.009"-diameter Pd wire was saturated in a pure $^4$He environment at 100 psi for 49.5 hours and subsequently evaporated, resulting in the release of only 21.7 picomoles of mass-4. This finding strongly suggests that the Mass-4 signal observed in Pd wires predominantly originates from $D_2$ rather than $^4$He. For Pd or other sample wires undergoing testing in novel nuclear experiments, it is essential to verify the absolute level of all mass-4 outgassed in these experiments against the initial amount present in the metals before the commencement of the experiment.

## 4. Conclusions

We have introduced a methodology development and conducted several studies concerning helium retention in materials subjected to various heating profiles and atmospheric concentrations. The assessment of trapped mass-4 in metals is notably crucial, as helium may be contained within materials like nylon, Teflon, and other technical materials utilized in the exploration of novel nuclear reactions. Without pretreatment, stainless steel 316 typically retains between 15 to 240 pmol of $^4$He, translating to an areal outgassing amount of 0.07 to 1.20 pmol/cm². Moreover, stainless steel may gradually reabsorb $^4$He from the atmosphere over time.

These studies underscore the necessity for pretreating all metals employed in the investigation of novel nuclear reactions to mitigate helium presence, particularly in experiments utilizing $^4$He presence as an indicator of reaction occurrence. Furthermore, our approach demonstrates a novel experimental method for assessing helium retention both before and after prolonged exposure to radiation from neutrons, alpha particles, and other sources with varying energies.

## 5. Acknowledgements

The authors would like to thank Kim Zinsmeyer, Chris Perez, Ruth Ogu, Robert Baca, Dr Peyton Thorn and Premitha Pansalawatte for their technical assistance and useful discussions. This work was supported by Department of Energy award No. DE-AR0001736, the Texas Research Incentive Program by Texas Tech University. The identification of commercial products, contractors, and suppliers within this article is for informational purposes only, and does not imply endorsement by Texas Tech University, their associates, or their collaborators.

## 6. Data Availability

The data that support the findings of this study are available within the supplementary materials and can be found on the TTU CEES website [19]. Additional materials will be provided from the corresponding author upon reasonable request.

## 8. Supplementary Materials

### 8.1 Additional Experimental Results

The leak rate and integrated leak was measured for several wire samples and metal types. The data in **Table 4** below contains the experimental results after the calibration using the NIST leak standard. The data in **Table 5** contain experimental results for additional technical materials.

**Table 4:** Total helium amount outgassed and release temperature of out-gassing from different materials.

| Control Experiments | Length of Wire [inch] | Total $^4$He Amount [pmol] | Average over multiple measurements [pmol] |
|---|---|---|---|
| W#1 | 6 | 3.8 | |
| W#2 | 6 | 7.4 | |
| W#3 | 6 | 0.8 | 4.0 ± 3.3 |

**Table 5:** Total helium amount outgassed and release temperature of out-gassing from different materials.

| Sample Name | Length of Wire [inch] | $^4$He Amount [pmol] | Subtracting Background from Control [pmol] | Notes |
|---|---|---|---|---|
| 24R304 SS insert with W #1 | 6 | 31.0 | 26.9±3.4 | 3" partially melted/evaporated |
| 24R304 SS insert with W #2 | 6 | 51.6 | 47.6±3.6 | 1.5" partially melted/evaporated |
| 24R304 SS insert with W #3 | 6 | 33.2 | 29.2±3.4 | 0.5" partially melted/evaporated |
| Safina wire wrap with W #1 | 6 | 22.6 | 18.6±3.3 | partially melted/evaporated, forming ball |
| Safina wire wrap with W #2 | 6 | 9.0 | 5.0±3.3 | partially melted/evaporated, forming ball |
| CFW wire wrap with W #1 | 6 | 15.2 | 11.2±3.3 | partially melted/evaporated |
| CFW wire wrap with W #2 | 6 | 9.8 | 5.8±3.3 | All melted / evaporated |
| 24R304 SS wrap with W | 6 | 20.8 | 16.8±3.3 | 0.5'' partially melted/evaporated |
| Safina wire wrap with W #5 | 6 | 9.6 | 5.6±3.3 | All melted / evaporated |

*

An additional example of the measured leak rate and integrated total leak are shown in **Figure 4**. These measurements were performed for a palladium wire sample.

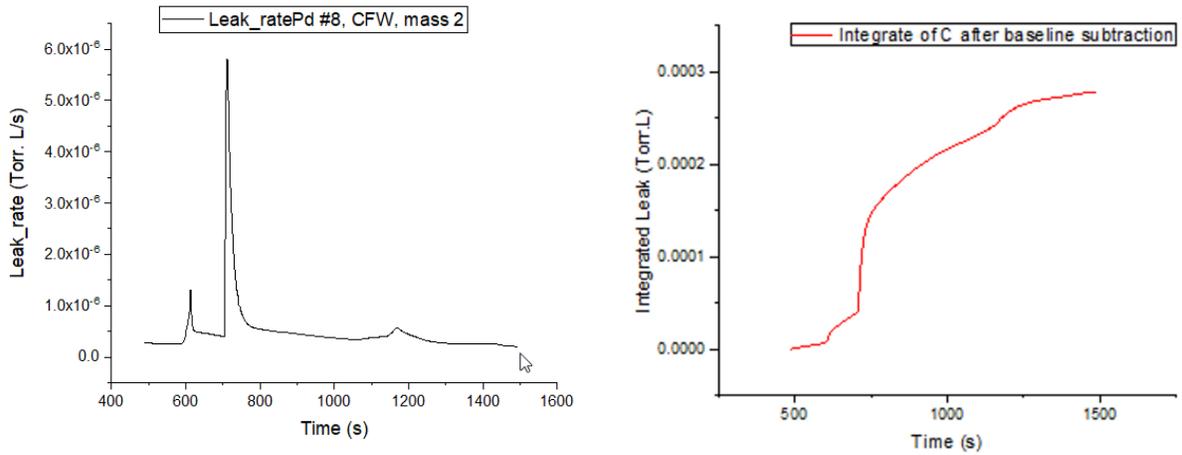

**Figure 4:** *Left*: The leak rate versus time for a palladium wire sample. *Right*: The integrated leak rate as a function of time for a palladium wire sample.